\documentclass[preprint,preprintnumbers, prd, floatfix,  superscriptaddress,nofootinbib] {revtex4-1}
\usepackage{graphicx}
\usepackage{epsfig}
\usepackage{bm}
\usepackage{amssymb}
\usepackage{float}
\usepackage{amsmath}
\usepackage{dcolumn}
\usepackage{cancel}
\usepackage[colorlinks]{hyperref}
\usepackage[usenames,dvipsnames]{color}
\usepackage{color}
\usepackage{epstopdf}
\usepackage{nccbbb}
\hypersetup{
     breaklinks=true,
    pdfstartview={FitH},    colorlinks=true,       
    linkcolor=blue,          
    citecolor=red,        
    filecolor=magenta,      
    urlcolor=blue,           
    anchorcolor=green,      
    linktocpage=true
}
\begin{document}

\title{A new pressure-parametric cosmological model}

\author{Yan-Hong Yao}
\email{yhy@mail.nankai.edu.cn}
\author{Xin-He Meng}
\email{xhm@nankai.edu.cn}

\affiliation{School of Physics, Nankai University, Tianjin 300071, China}

\begin{abstract}
We put forward a pressure-parametric model to study the tiny deviation from cosmological constant(CC) behavior of the dark sector accelerating the expansion of the Universe. Data from cosmic microwave background (CMB) anisotropies, baryonic acoustic oscillations (BAO), Type Ia supernovae (SN Ia) observation are applied to constrict the model parameters. The constraint results show that such model suffers with $H_0$ tension as well. To realize this model more physically, we reconstruct it with the quintessence and phantom scalar fields, and find out that although the model predicts a quintessence-induced acceleration of the Universe at past and present, at some moment of the future, dark energy's density have a disposition to increase.

\end{abstract}

\maketitle

\section{Introduction}
\label{intro}
After the discovery in 1998 from the observations of nearby Type Ia supernovae\cite{riess1998,perlmutter1999measurements}, ever-growing observation, such as  Baryon Oscillation Spectroscopic Survey\cite{Schlegel2009The}, the Dark Energy Survey\cite{Crocce2016Galaxy}, the Planck\cite{Collaboration2015Planck} and some other observations have confirmed the accelerated expansion of the Universe in recent time.

The $\Lambda$CDM model is a simple and valid cosmological model, in which cosmological constant $\Lambda$ plays the role of the invisible fuel, called as dark energy, that accelerates the expansion of the Universe. Although the $\Lambda$CDM model provides an good fit to almost all of the cosmological data so far, such model is theoretical problematic for its two unsolved problem, i.e., the fine-tuning problem and the coincidence problem, the fine-tuning problem indicates the huge disagreement between the theoretical vacuum density and the observational one, while the coincidence problem questions why is the observational vacuum density coincidentally comparable with the dark matter density at the present epoch of the Universe. To solve these two problem, other alternative model have been put forward, including scalar field models such as quintessence\cite{Caldwell1998Cosmological}, phantom\cite{Caldwell2002A}, dilatonic\cite{Piazza2004Dilatonic}, tachyon\cite{Padmanabhan2002Accelerated}and quintom\cite{Bo2006Oscillating} etc. and modified gravity models such as braneworlds \cite{maartens2010brane}, scalar-tensor gravity \cite{esposito2001scalar}, higher-order gravitational theories \cite{capozziello2005reconciling,das2006curvature}.

Besides constructing fundamental models like scalar field models and modified gravity models mentioned above, parametrization is another useful way for us to characterize the properties of dark energy. Although up till the present moment, a majority of cosmologists favor the equation of state (EoS) parametrization rather than other kind of parametrization, we focus on late-time total pressure parametrization in this paper. Pressure parametrization is first suggested by A. A. Sen in 2008\cite{Sen2008Deviation}, in his work, A. A. Sen proposed a parametric model of dark energy, which is the Taylor series expansion around the CC behavior $p(a)=-\Lambda+p_{1}(1-a)+\frac{1}{2}p_{2}(1-a)^{2}+...$ , to study the small deviation from the $\Lambda$CDM model, here $a$ is the scale factor. Seven years later, Q. Zhang et al.\cite{Zhang2015Exploring} suggested two other pressure-parametric scenario for dark energy, namely $p(z)=p_{a}+p_{b}z$ and $p(z)=p_{a}+\frac{p_{b}}{1+z}$. It turn out that both models detect subtle hints $\omega_{de}<-1$ from the Type Ia supernovae and BAO data, which indicates possibly a phantom dark energy scenario at present. Follow this theoretical line, D. Wang et al.\cite{Wang2017A} raised a unified dark fluid model that parameterize the total pressure as $p(z)=p_{a}+p_{b}(z+\frac{z}{1+z})$. Later on, J. C. Wang et al.\cite{junchaowang2018A} proposed another pressure-parametric model, in which the total pressure is parameterized as $p(z)=p_{a}+p_{b}$ln$(1+z)$. To continue study the deviation from CC behavior of dark energy, we propose a new pressure-parametric dark energy model, place it with data from  CMB anisotropies, BAO, SN Ia observations, recover it in the scenarios of quintessence and phantom fields, and explore the behaviors of the scalar field and potential in the following of the paper.

The contents of this paper are as follows. In Section 2, we present a pressure-parametric model. In Section 3, we confront the new model with the data from  CMB anisotropies, BAO, SN Ia observations. In Section 4, we reconstruct the quintessence and phantom fields from the model and observations and investigate the evolution behaviors of the scalar field and potential. A brief conclusion of the paper are presented at the final section.
We set the light velocity $c=1$, as well as km/s/Mpc=1.

\section{ The new pressure-parametric model}
\label{sec:2}
Assuming that the Einstein's equation correctly describes the evolution behavior of space-time on cosmological scale, a homogeneous, isotropic, and spatially flat cosmological model is constituted by the Friedmann equation
\begin{equation}\label{1}
  H^{2}=\frac{\kappa^{2}}{3}\rho,
\end{equation}
the continuity equation
\begin{equation}\label{2}
  \dot{\rho}+3H(\rho+p)=0,
\end{equation}
and the EoS $\omega=\frac{p}{\rho}$. Here $H$ is the Hubble parameter, $\kappa=\sqrt{8\pi G}$ is the inverse of the reduced Planck mass, $\rho$ and $p$ denote the energy density and pressure respectively.
As is mentioned in the introduction, we specify a model by parameterizing the total pressure rather than the EoS. In fact, these two seemingly different ways are equivalent, because we can obtain the $p(z)-z$ relation by inserting the EoS into the continuity equation and  recover the EoS by inserting the $p(z)-z$ relation into the continuity equation.

The $p(z)-z$ relation chosen to determine our pressure-parametric model is
\begin{equation}\label{3}
  p=p_{a}+p_{b}\frac{1+z^{3}}{1+z},
\end{equation}
where $p_{a}$ and $p_{b}$ are two constant.
From (3), one notes that as $z\rightarrow-1$, $p\rightarrow p_{a}+3p_{b}$, so the pressure is a finite value when we consider $a\rightarrow\infty$ (Since $a=\frac{a_{0}}{1+z}=\frac{1}{1+z}$). This result is different with the conclusion deduced from the $p(z)-z$ relations $p(z)=p_{a}+\frac{p_{b}}{1+z}$\cite{Zhang2015Exploring}, $p(z)=p_{a}+p_{b}(z+\frac{z}{1+z})$\cite{Wang2017A} and $p(z)=p_{a}+p_{b}$ln$(1+z)$\cite{junchaowang2018A}, because space-times described by all of these models encounter little rip.
Inserting the $p(z)-z$ relation into the equation of energy conservation and solving the equation leads to
\begin{equation}\label{4}
  \rho(a)=-( p_{a}+3p_{b})+p_{b}(-3a^{-2}+\frac{9}{2}a^{-1})+Ca^{-3}.
\end{equation}
$C=\rho_{0}+p_{a}+\frac{3}{2}p_{b}$ is an integration constant, $\rho_{0}$ is the present energy density.  The first
term of the right side of (4) is a constant term therefore can be interpreted as the CC term, the last term is proportional to $a$ to the power of minus three so can be regarded as the density of the usual matter and dark matter, the second term, however, characterize the small deviation from the CC behavior of dark energy.
For the sake of simplicity, we redefine two dimensionless parameters: $\Omega_{m0}=\frac{\rho_{0}+p_{a}+\frac{3}{2}p_{b}}{\rho_{0}}$
(the present dimensionless matter parameter) and $\eta=\frac{3p_{b}}{2\rho_{0}}$, then the late-time total energy density and pressure can be expressed as:
\begin{eqnarray}
  \rho(a)&=&\rho_{0}(1-\Omega_{m0}+\eta(-2a^{-2}+3a^{-1}-1)+\Omega_{m0}a^{-3}),\\
   p(a) &=& \rho_{0}(-1+\Omega_{m0}+\eta(\frac{2}{3}a^{-2}-2a^{-1}+1)).
\end{eqnarray}
According to the Friedmann equation (1), we can derive
the dimensionless Hubble parameter:
\begin{equation}\label{7}
  E^{2}=\frac{H^{2}}{H_{0}^{2}}=1-\Omega_{m0}+\eta(-2a^{-2}+3a^{-1}-1)+\Omega_{m0}a^{-3}.
\end{equation}
Since we will confront the model with the observations in the next section, it is convenient to use redshift $z$ as variable instead of scale factor $a$. Also, some of observational data to be matter with physical processes happened in the early epoch of the Universe when radiation still played a important role to the evolution of the cosmos, therefore we rewriting the equation (7) by introducing a radiation term, i.e.
\begin{equation}\label{7}
  E^{2}=1-\Omega_{m0}-\Omega_{r0}+\eta(-2(1+z)^{2}+3(1+z)-1)+\Omega_{m0}(1+z)^{3}+\Omega_{r0}(1+z)^{4},
\end{equation}
where $\Omega_{r0}=\omega_{\gamma}(1+\frac{7}{8}(\frac{4}{11})^{\frac{4}{3}}N_{eff}) h^{-2}, \omega_{\gamma}=2.47 \times 10^{-5}, N_{eff}=3,  h=\frac{H_{0}}{100}$.  From the (8), one finds that the model can be thought as a Taylor expansion of $E^{2}(z)$, in fact, it can be thought as a model with matter, radiation, curvature, CC term and a fluid which in connect with the curvature, its EoS is $-\frac{2}{3}$. In this sense, the Universe is close when $\eta>0$ and open when $\eta<0$.

\section{Confront the model with observation}
\label{sec:3}
In this section, we confront the model with CMB, BAO, SN Ia observational data.
\subsection{CMB measurements}
The observed angular scale of CMB fluctuations are mainly determined by the sound horizon at decoupling $r_{s}(z_{\ast})$, and the angular distance at decoupling $d_{A}(z_{\ast})$. $r_{s}(z_{\ast})$ depends on the dominators in the energy budget at $z>z_{\ast}$ , and $d_{A}(z_{\ast})$ depends on the dominator in the energy budget at $z>z_{\ast}$. Where  $r_{s}(z_{\ast})$ and $d_{A}(z_{\ast})$ are showed by \cite{Efstathiou2010Cosmic}
\begin{eqnarray}
  r_{s}(z_{\ast})&=&\frac{1}{\sqrt{3}}\int_{0}^{\frac{1}{1+z_{\ast}}}\frac{da}{a^{2}H(a)\sqrt{1+\frac{3\omega_{b}a}{4\omega_{\gamma}}}},\\
  d_{A}(z_{\ast})&=&\frac{1}{1+z}\int_{0}^{z_{\ast}}\frac{d\tilde{z}}{H}.
   \end{eqnarray}
The redshift at decoupling $z_{\ast}$ is given by\cite{Hu1996Small}
\begin{eqnarray}
  z_{*}&=&1048(1+0.00124\omega_{b}^{-0.738})(1+g_{1}\omega_{m}^{g_{2}}),\\
   g_{1}&=&\frac{0.0783\omega_{b}^{-0.238}}{1+39.5\omega_{b}^{0.763}},\\
  g_{2}&=&\frac{0.56}{1+21.1\omega_{b}^{1.81}},
\end{eqnarray}
where $\omega_{m}=\Omega_{m0}h^{2},\omega_{b}=\Omega_{b0}h^{2}$.

Most of the information contained in the CMB power spectrum can be compressed into two shift parameters. The first, $R$, is defined as
\begin{equation}
  R=\sqrt{\Omega_{m}H_0^2}(1+z_{\ast})d_{A}(z_{\ast}),
\end{equation}
and the second, $l_{A}$, is given by
\begin{equation}
  l_{A}=\pi(1+z_{\ast})\frac{d_{A}(z_{\ast})}{r_{s}(z_{\ast})}.
\end{equation}

In this work, we use the Planck 2018 compressed likelihood\cite{Chen2018Distance} with these two shift parameters. The value for $(R,l_{A},\omega_{b})=(1.7502,301.471,0.02236)$ with errors $(0.0046,0.09,0.00015)$, and
the covariance is
\begin{equation}
  D_{CMB}=  \left(
      \begin{array}{ccc}
        1 & 0.46 &-0.66 \\
       0.46 & 1 & -0.33 \\
       -0.66&-0.33& 1\\
      \end{array}
    \right),
\end{equation}
so that the elements of the covariance matrix $C_{ij} =\sigma _i \sigma_j D_{ij}$.
then we have
\begin{eqnarray}
  \chi_{CMB}^{2}&=&s^{T} C_{CMB}^{-1} s,\\
  s^{T} &=& (R-1.7502,l_{A}-301.471,\omega_{b}-0.02236).
   \end{eqnarray}

\subsection{Baryon acoustic oscillations}
The relative BAO distance is defined as
\begin{equation}
   d_{z}(z)=\frac{r_{s}(z_{d})}{D_{V}(z)},
\end{equation}
with $D_{V}(z)=[(1+z)^{2}d_{A}(z)^{2}\frac{z}{H(z)}]^{\frac{1}{3}}$. The drag epoch $z_{d}$ is the epoch when baryons are released from the Compton drag of the photons. It can be calculated using\cite{Eisenstein1997Baryonic}
  \begin{eqnarray}
     z_{d}&=&1291\frac{\omega_{m}^{0.251}}{1+0.659\omega_{m}^{0.828}}(1+b_{1}\omega_{b}^{b_{2}}),\\
     b_{1}&=&0.313\omega_{m}^{-0.419}(1+0.607\omega_{m}^{0.674}),\\
     b_{2}&=&0.238\omega_{m}^{0.223}.
   \end{eqnarray}
   We follow \cite{Collaboration2015Planck} and use four measurements from 6dFGS at $z_{eff}$ = 0.106, the recent SDSS
main galaxy (MGS) at $z_{eff}$ = 0.15 of \cite{Ross2014The} and $z_{eff}$= 0.32 and 0.57 for the Baryon Oscillation
Spectroscopic Survey (BOSS)\cite{Anderson2013The}. We consider the $\chi_{BAO}^2$ of the form as:
   \begin{eqnarray}
     \chi_{BAO}^{2}&=&t^{T} C_{BAO}^{-1} t,
   \end{eqnarray}
with
\begin{eqnarray}
  t^{T} &=&(d_{z}(0.106)-0.336,d_{z}(0.15)-0.2239,d_{z}(0.32)-0.1181,d_{z}(0.57)-0.07206) ,\\
     C_{BAO}^{-1} &=& diag(4444.44,14071.64,183411.36,2005139.41).
\end{eqnarray}

\subsection{Type Ia supernovae}
We employ the Joint Light-curve Analysis (JLA) supernova sample \cite{betoule2014improved}, so the distance modulus is assumed as the following formula
\begin{equation}
  \mu_{obs}=m_B-(M_B-\alpha x_1+\beta c),
\end{equation}
where $m_B$ and $M_B$ are SN Ia peak apparent magnitude and SN Ia absolute magnitude respectively, $\alpha$ and $\beta$ are two constants,
$c$ is the color parameter and $ x_1$ is the stretch factor. Now,
\begin{equation}
\chi_{SN}^2=\Delta^{T}C_{SN}^{-1}\Delta,
\end{equation}
where $\Delta= \mu-\mu_{obs}$ and $C_{SN}$ is the covariance matrix.

Finally, the $\chi^{2}$ is given by
 \begin{equation}
   \chi^2=\chi_{CMB}^2+\chi_{BAO}^2+\chi_{SN}^2.
 \end{equation}

\begin{figure}
  \includegraphics[width=0.55\textwidth]{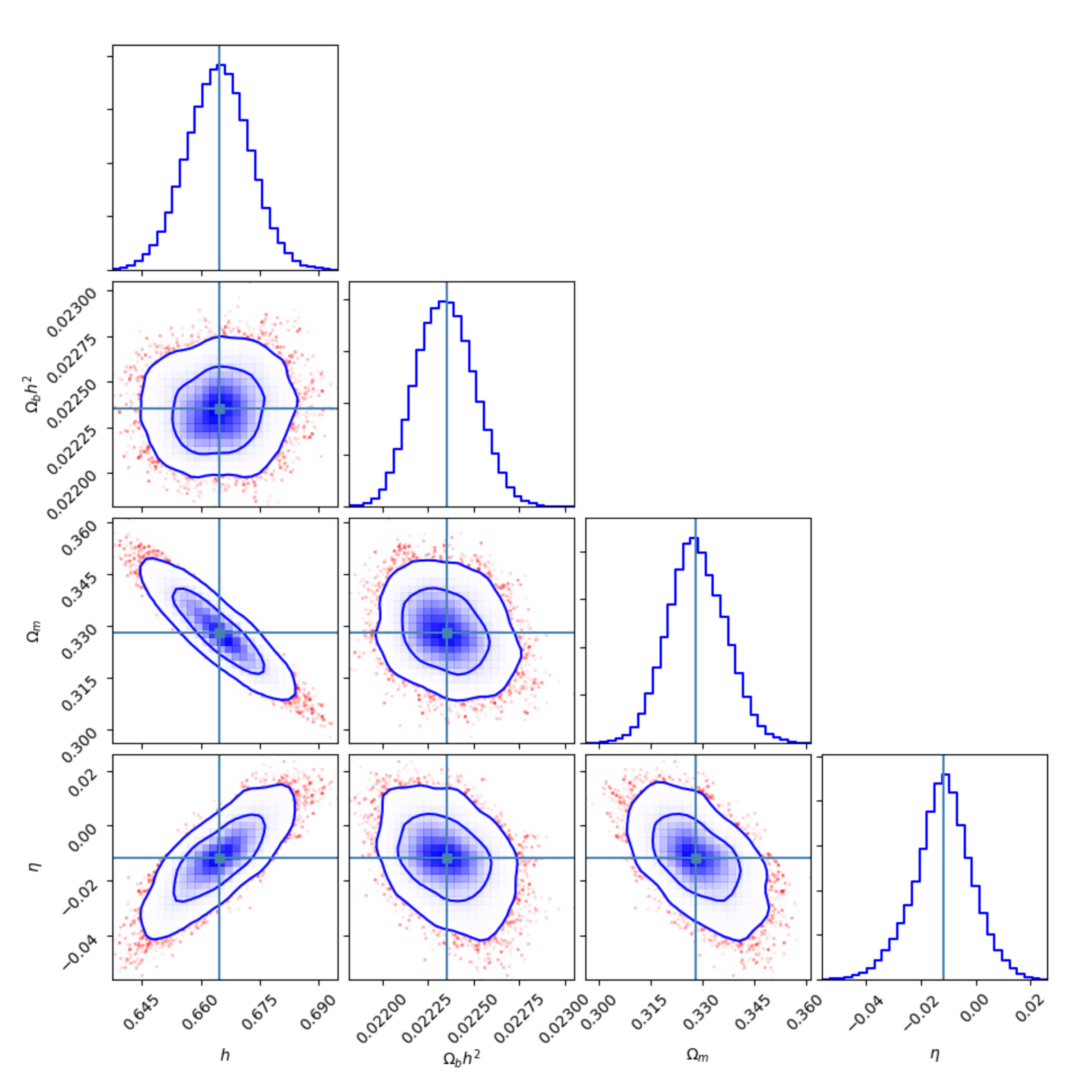}
\caption{$1\sigma$ and $2\sigma$ confidence regions for parameters.}
\label{fig:1}       
\end{figure}

\begin{table}
\begin{center}
\begin{tabular}{|c|c|c|}
\hline  & the new model   & the $\Lambda$CDM model
\\ \hline
$\Omega_{m0}$    & $0.328 _{-0.008}^{+0.009}$  & $0.322 _{-0.006}^{+0.007}$
                     \\
$\eta$         &  $-0.012_{-0.01}^{+0.01}$  &   -
                     \\
 $h$          &    $ 0.665_{-0.008}^{+0.008}$  & $ 0.672_{-0.004}^{+0.005}$
                       \\
 $\omega_{b}$          &    $ 0.0223_{-0.0002}^{+0.0002}$ & $ 0.0223_{-0.0001}^{+0.0001}$
                       \\  \hline
 $\chi_{min}^2$    &   686.677 &  687.847
                         \\
 $AIC$           &   694.677   & 693.847
                       \\
\hline
\end{tabular}
\caption{$\chi_{min}^2$, $AIC$ and 1$\sigma$ confidence interval of the parameters for the new model and the $\Lambda$CDM model }
\label{tab:2}
\end{center}
\end{table}

The observational constraints on the model parameters $\Omega_{m0}, \eta, h, \omega_{b}$ are presented in Fig.\ref{fig:1}, Table\ref{tab:2}. In Table\ref{tab:2}, we also add some fitting results for parameters in the $\Lambda$CDM model by using the same data sets. As we know, the local measurement $H_{0}=74.03\pm1.42$  from Riess\cite{Riess2019A} exhibits a strong tension with the Planck 2018 release \cite{Collaboration2018Planck} $H_{0} = 67.4\pm0.5$ at the 4.4$\sigma$ level. Table\ref{tab:2} shows that our best-fit value of $H_{0}$ in the new model is below the measured value from the Planck 2018's result or the value in the $\Lambda$CDM model by using the same data sets. One can be told that this is reasonable from the Fig.\ref{fig:1}, which shows that $\eta$ is positively correlated with $h$, as a result, a tiny minus values of the parameter $\eta$ leads to a smaller values of Hubble constant $H_0$. Therefore, such model suffers with $H_0$ tension as well.
 We also consider the Akaike information criterion (AIC) to compare the new model and the $\Lambda$CDM model. We have AIC = $\chi_{min}^2 + 2k$, where $k$ denotes the number of cosmological parameters. In fact, we only care about the relative value of the AIC
between two different models, i.e., $\Delta$ AIC = $\Delta\chi_{min}^2 + 2\Delta k$. A model with a smaller value of AIC
is a more supported model. In this paper, the $\Lambda$CDM model serves as
a reference model. From Table\ref{tab:2}, we have $\Delta AIC=0.83$.

\section{The reconstruction}
\begin{figure}
  \includegraphics[width=0.5\textwidth]{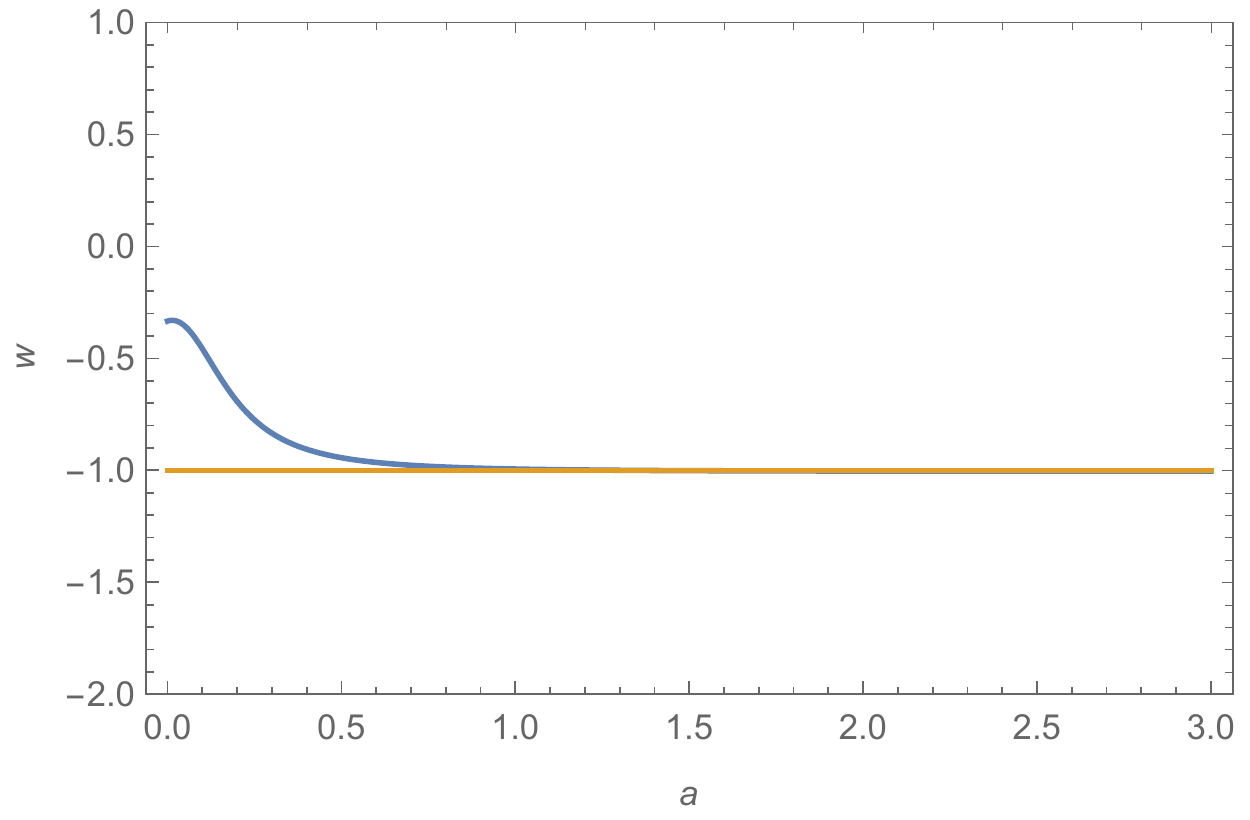}
\caption{EoS for dark energy $w$ versus scale factor $a$, blue line corresponds to the new model, orange line corresponds to the $\Lambda$CDM model.}
\label{fig:2}
\end{figure}

To realize the new model more physically, in this section we reconstruct it with scalar fields. Firstly, we need to consider the EoS for dark energy in this model that is allowed by the data. From (5) and (6), we obtain
\begin{equation}
  w=\frac{p}{\rho}=\frac{-1+\Omega_{m0}+\eta(\frac{2}{3}a^{-2}-2a^{-1}+1)}{1-\Omega_{m0}+\eta(-2a^{-2}+3a^{-1}-1)}.
\end{equation}
By substituting the best-fit values for $\Omega_{m0}$, $\eta$ and plotting the relation $w(a)-a$ on the Fig.\ref{fig:2},
 one can infer that the equation of state for dark energy in the new model is larger than $-1$ if $a<\frac{4}{3}$ and smaller than $-1$ if $a>\frac{4}{3}$. Therefore, it needs both quintessence and phantom to reconstruct the model.

As a result, let us start the reconstruction by giving the action for both quintessence and phantom as
\begin{equation}
  S_{\phi}=-\int d^{4}x\sqrt{-g}[\frac{b}{2}\partial_{\mu}\phi\partial^{\mu}\phi+V(\phi)].
\end{equation}
Here $\frac{b}{2}\partial_{\mu}\phi\partial^{\mu}\phi$ denotes the kinetic term while $V(\phi)$ potential term. $b=1,-1$ corresponding to the case
of quintessence and phantom scalar respectively.
From the stress energy tensor $T_{\mu\nu}=-\frac{2}{\sqrt{-g}}\frac{\delta S_{\phi}}{\delta g^{\mu\nu}}$, assuming a homogeneous, isotropic space-time, we have
\begin{eqnarray}
  \rho_{\phi} &=& \frac{b}{2}\dot{\phi}^2+V(\phi) ,\\
  p_{\phi} &=& \frac{b}{2}\dot{\phi}^2-V(\phi).
\end{eqnarray}
For simplicity, we consider that the Universe is nonradiative in the rest of the section.
\subsection{The quintessence case}
Regarding the quintessence as dark energy, from (5),(6),(31) and (32) one obtains
\begin{eqnarray}
 \rho &=& \frac{1}{2}\dot{\phi}^2+V(\phi)=\rho_{0}(1-\Omega_{m0}+\eta(-2a^{-2}+3a^{-1}-1)),\\
  p &=& \frac{1}{2}\dot{\phi}^2-V(\phi)=\rho_{0}(-1+\Omega_{m0}+\eta(\frac{2}{3}a^{-2}-2a^{-1}+1)).
\end{eqnarray}
Deriving kinetic term and potential term from (33) and (34), leads to
\begin{eqnarray}
  \dot{\phi}^2 &=& \rho_{0}\eta(-\frac{4}{3}a^{-2}+a^{-1}), \\
  V(\phi) &=& \rho_{0}(1-\Omega_{m0}+\eta(-\frac{4}{3}a^{-2}+\frac{5}{2}a^{-1}-1)),\\
  \frac{d\phi}{da}&=&\pm\frac{1}{\kappa a}\sqrt{\frac{\eta(-4a^{-2}+3a^{-1}) }{(1-\Omega_{m0}+\eta(-2a^{-2}+3a^{-1}-1)+\Omega_{m0}a^{-3})}}.
\end{eqnarray}
As we can see, assuming $\eta<0$, a real scalar field condition requires $a<\frac{4}{3}$, which is also the requirement for $\omega_{\phi}>-1$
\begin{figure}[h]
\begin{minipage}{0.45\linewidth}
  \centerline{\includegraphics[width=1\textwidth]{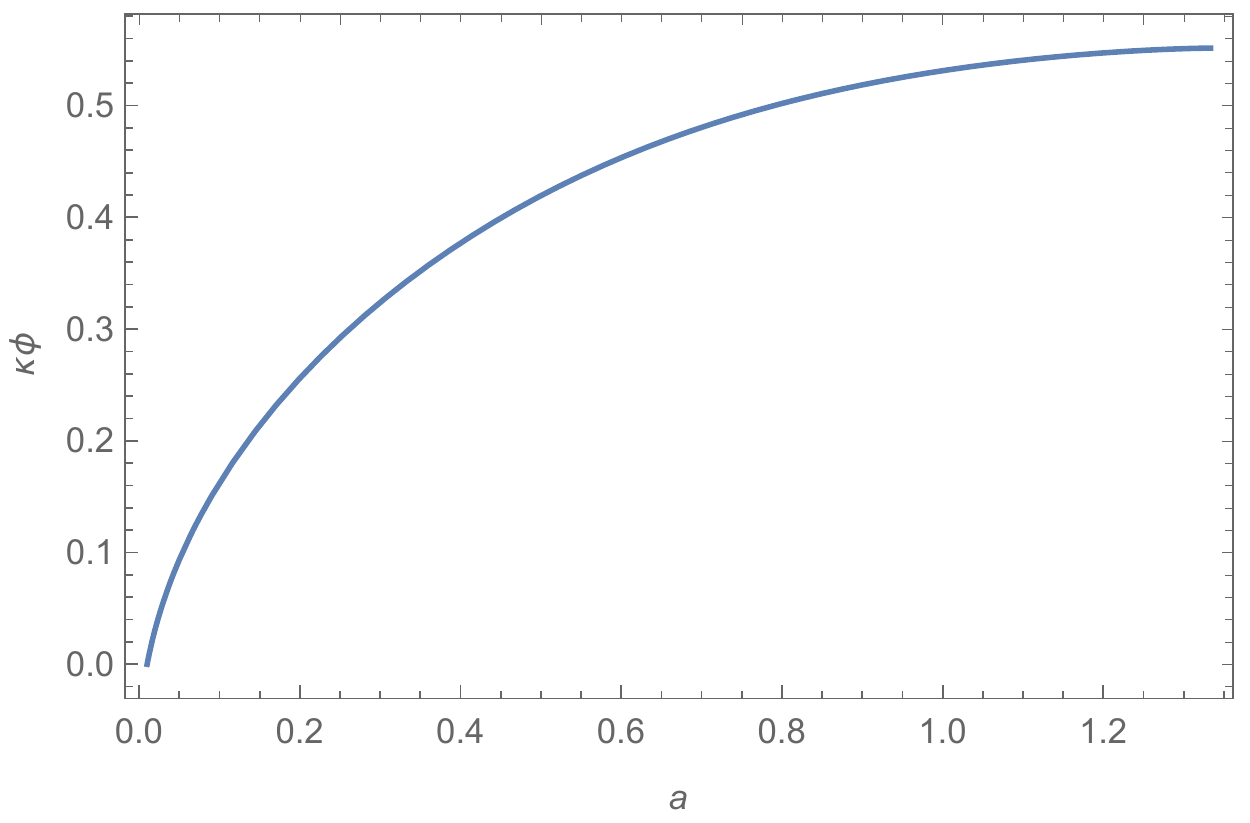}}
  \label{a}
\end{minipage}
\begin{minipage}{0.45\linewidth}
  \centerline{\includegraphics[width=1\textwidth]{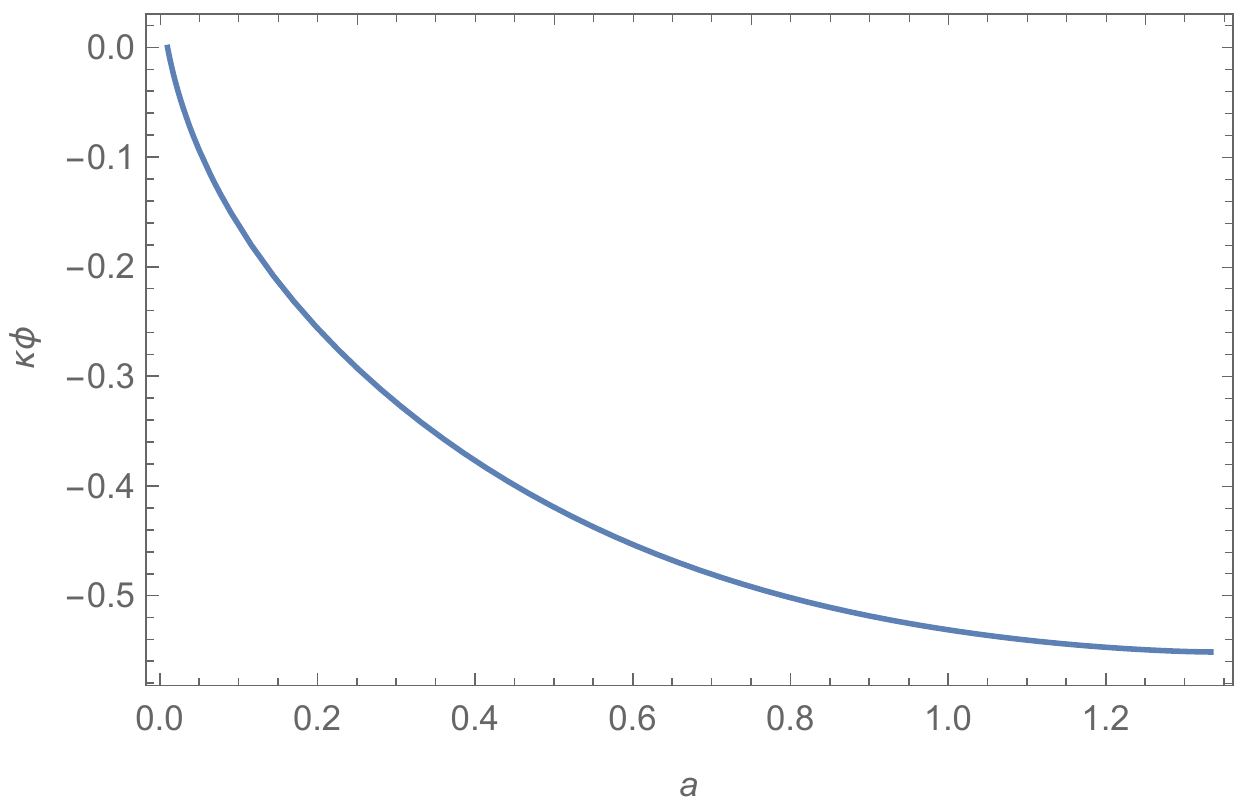}}
  \label{a}
\end{minipage}
\begin{minipage}{0.45\linewidth}
  \centerline{\includegraphics[width=1\textwidth]{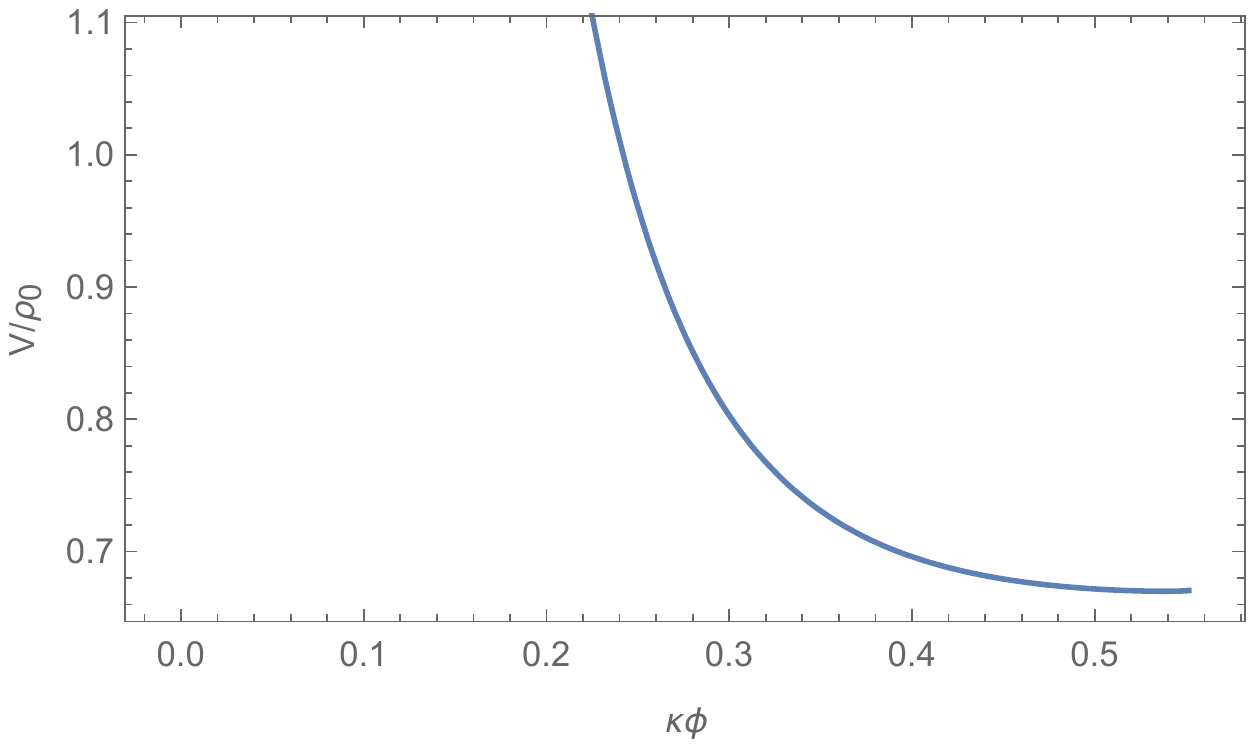}}
  \label{a}
\end{minipage}
\begin{minipage}{0.45\linewidth}
  \centerline{\includegraphics[width=1\textwidth]{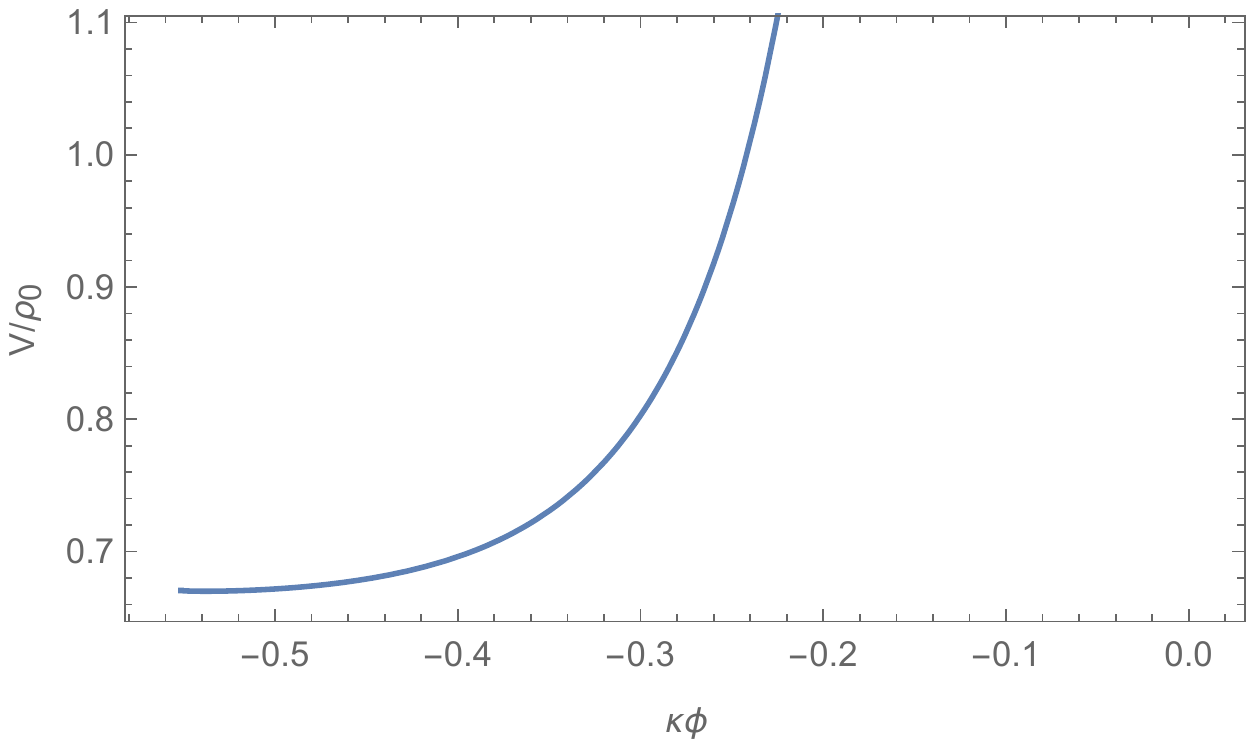}}
  \label{a}
\end{minipage}
\caption{ Upper panels of the quintessence field $\kappa\phi$ versus scale factor $a$, and lower panels of the potential $V/\rho_{0}$ versus quintessence field $\kappa\phi$ versus are shown. Panels on the left corresponding to plus while panels on the right corresponding to minus in (37). We have use $\Omega_{m0}=0.328,\eta=-0.012$ and $\phi(0.01)=0$ here.}
\label{fig:3}
\end{figure}

We plot the $\phi(a)$ and $V(\phi)$ in the Fig.\ref{fig:2} by numerically solving the (36) and (37) with $\Omega_{m0}=0.328,\eta=-0.012$ and $\phi(0.01)=0$. We can see that the quintessence field $\kappa\phi$ increases monotonically with the increasing $a$ on the left and decreases monotonically with the increasing $a$ on the right, while the potential $V/\rho_{0}$ decreases monotonically with the increasing quintessence field $\kappa\phi$ on the left and increases monotonically with the increasing quintessence field $\kappa\phi$ on the right.

\subsection{The phantom case}
Regarding the phantom as dark energy, from (5),(6),(31) and (32) one obtains
\begin{eqnarray}
 \rho &=& -\frac{1}{2}\dot{\phi}^2+V(\phi)=\rho_{0}(1-\Omega_{m0}+\eta(-2a^{-2}+3a^{-1}-1)),\\
  p &=& -\frac{1}{2}\dot{\phi}^2-V(\phi)=\rho_{0}(-1+\Omega_{m0}+\eta(\frac{2}{3}a^{-2}-2a^{-1}+1)).
\end{eqnarray}
Deriving kinetic term and potential term from (38) and (39), leads to
\begin{eqnarray}
  \dot{\phi}^2 &=& -\rho_{0}\eta(-\frac{4}{3}a^{-2}+a^{-1}), \\
  V(\phi) &=& \rho_{0}(1-\Omega_{m0}+\eta(-\frac{4}{3}a^{-2}+\frac{5}{2}a^{-1}-1)),\\
  \frac{d\phi}{da}&=&\pm\frac{1}{\kappa a}\sqrt{\frac{-\eta(-4a^{-2}+3a^{-1}) }{(1-\Omega_{m0}+\eta(-2a^{-2}+3a^{-1}-1)+\Omega_{m0}a^{-3})}}.
\end{eqnarray}
Again, assuming $\eta<0$, a real phantom scalar field condition requires $a>\frac{4}{3}$.

\begin{figure}[h]
\begin{minipage}{0.45\linewidth}
  \centerline{\includegraphics[width=1\textwidth]{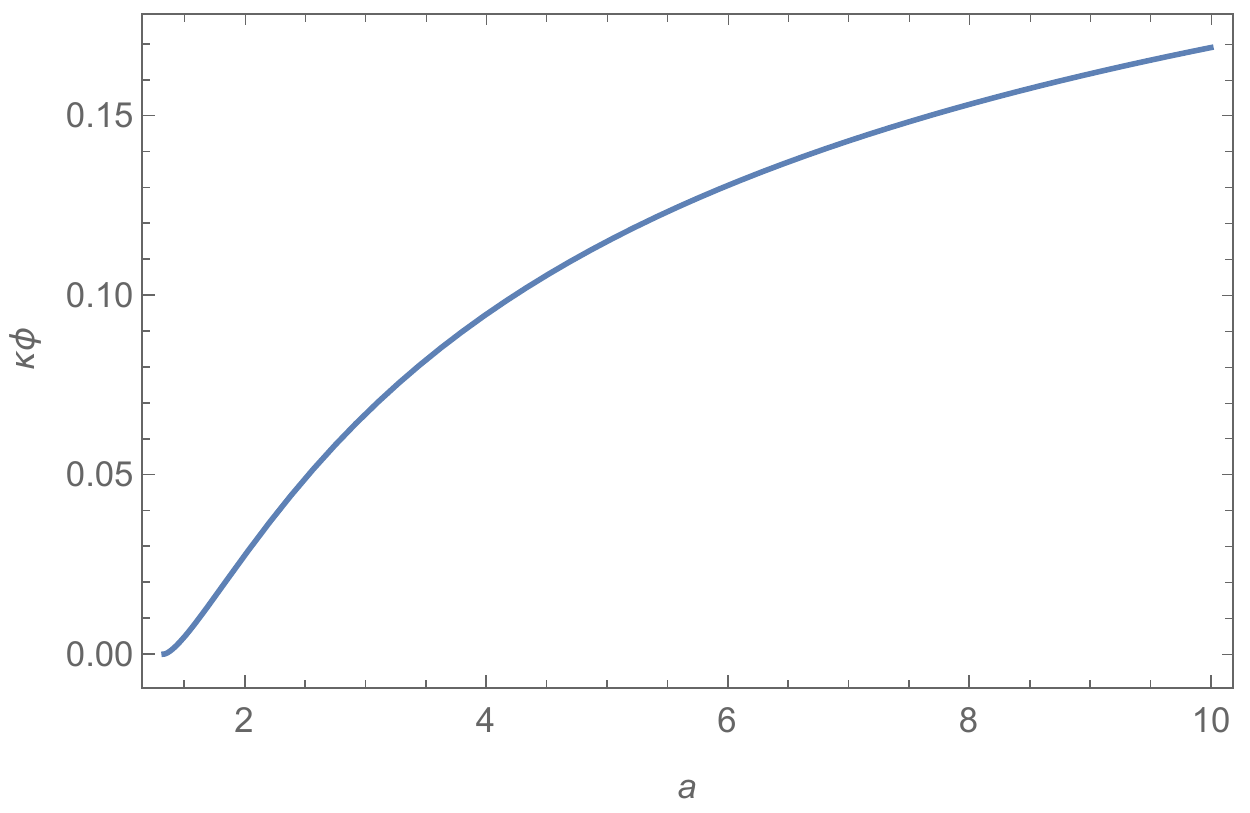}}
  \label{a}
\end{minipage}
\begin{minipage}{0.45\linewidth}
  \centerline{\includegraphics[width=1\textwidth]{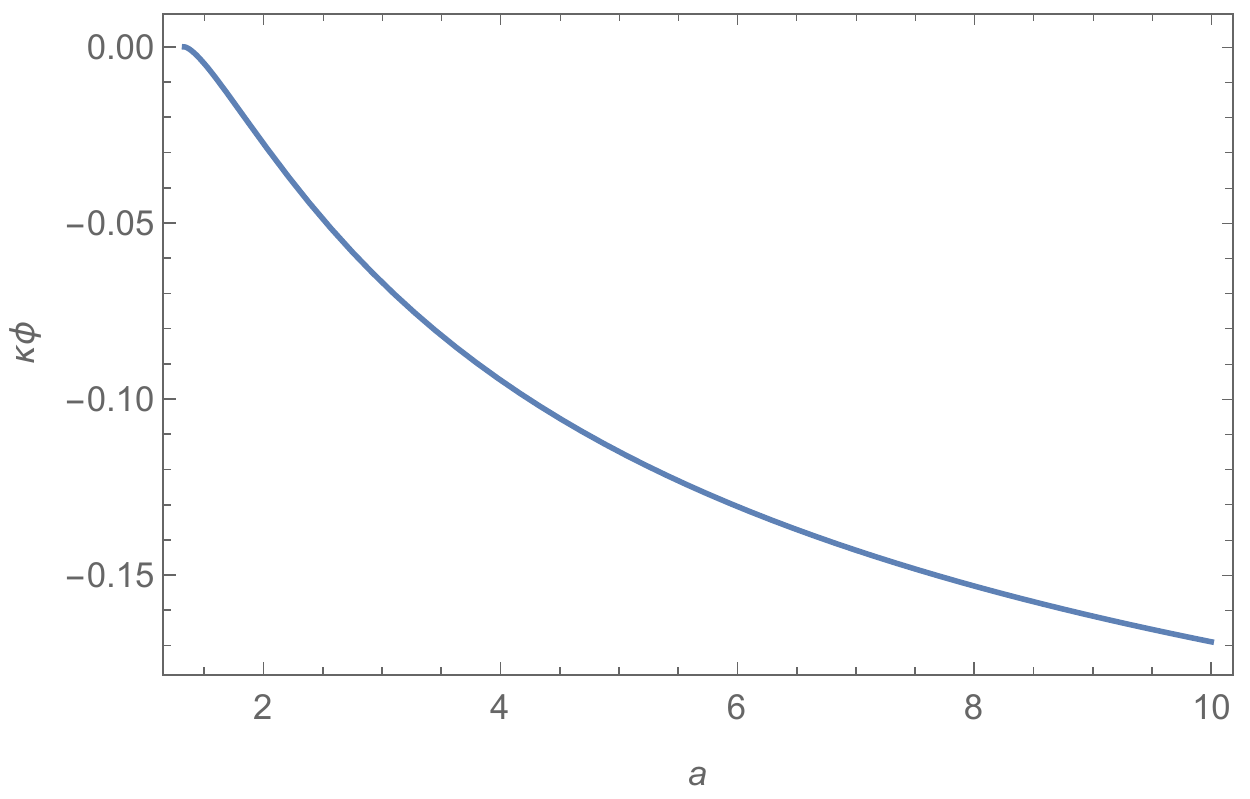}}
  \label{a}
\end{minipage}
\begin{minipage}{0.45\linewidth}
  \centerline{\includegraphics[width=1\textwidth]{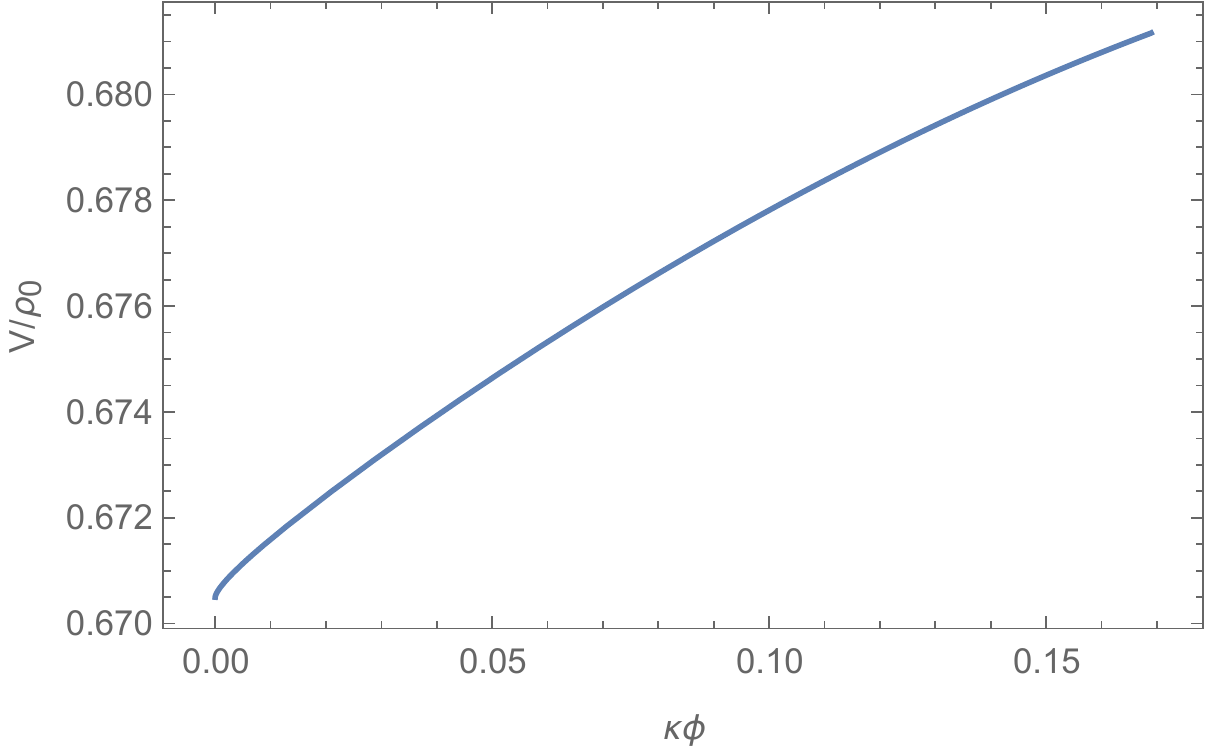}}
  \label{a}
\end{minipage}
\begin{minipage}{0.45\linewidth}
  \centerline{\includegraphics[width=1\textwidth]{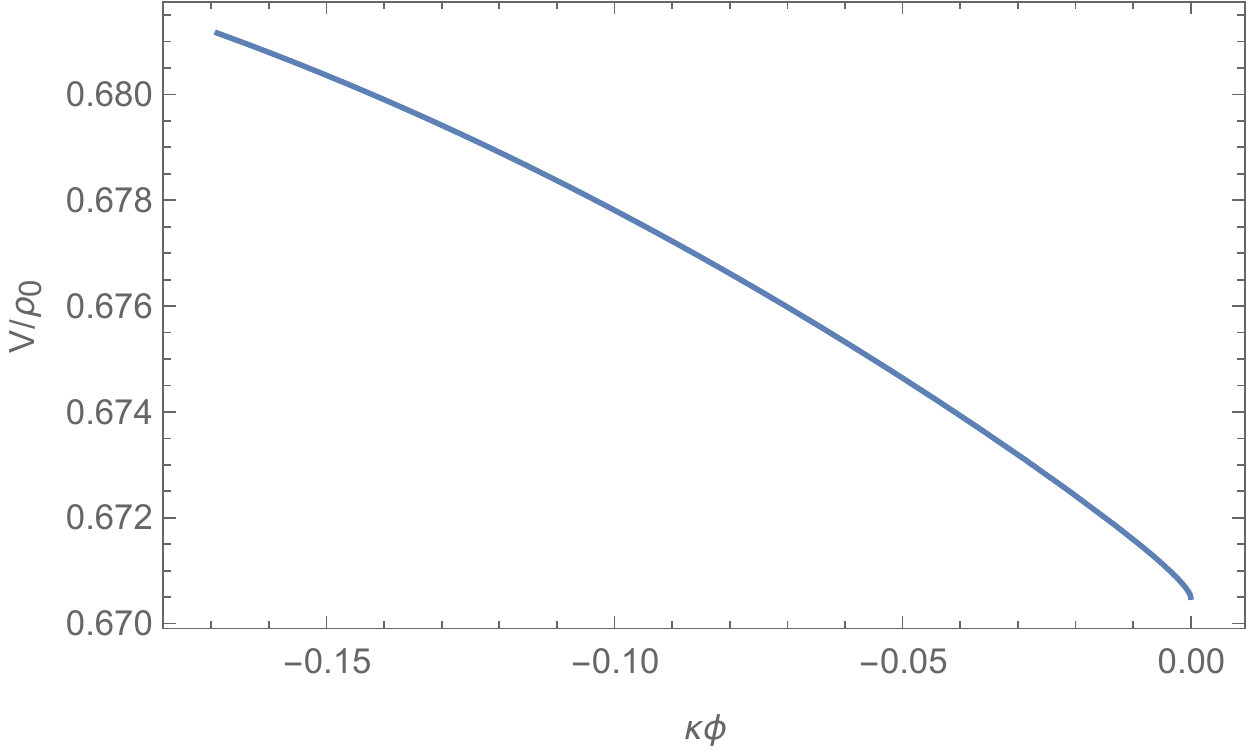}}
  \label{a}
\end{minipage}
\caption{ Upper panels of the phantom field $\kappa\phi$ versus scale factor $a$, and lower panels of the potential $V/\rho_{0}$ versus phantom field $\kappa\phi$ versus are shown. Panels on the left corresponding to plus while panels on the right corresponding to minus in (42). We have use $\Omega_{m0}=0.328,\eta=-0.012$ and $\phi(0.01)=0$ here.}
\label{fig:4}
\end{figure}

We plot the $\phi(a)$ and $V(\phi)$ in the Fig.\ref{fig:3} by numerically solving the (41) and (42) with $\Omega_{m0}=0.328,\eta=-0.012$ and $\phi(0.01)=0$.
We can see that the quintessence field $\kappa\phi$ increases monotonically with the increasing $a$ on the left and decreases monotonically with the increasing $a$ on the right, while the potential $V/\rho_{0}$ increases monotonically with the increasing quintessence field $\kappa\phi$ on the left and decreases monotonically with the increasing quintessence field $\kappa\phi$ on the right.

\subsection{The evolution of the scalar field}
Here we only discuss the plus case, the minus case is similar.
From $(35)-(37),(40)-(42)$ and Fig.\ref{fig:2} ,Fig.\ref{fig:3}, we know that the scalar rolling begins at $\dot{\phi}(0.01)=\sqrt{158.8\rho_{0}}, V(a=\hat{a})=157.684\rho_{0}$.
At this period, the scalar behaves like a quintessence, therefore the equation of evolution governed the field is
\begin{equation}
   \ddot{\phi}+3H\dot{\phi}+V'(\phi)=0,
\end{equation}
so the field rolls down to the bottom of the potential due to the initial kinetic energy and the force induced by the potential, when the quintessence field reach the bottom of the potential, it turn into a phantom field.

After then, according to the equation of evolution
\begin{equation}
   \ddot{\phi}+3H\dot{\phi}-V'(\phi)=0,
\end{equation}
the field rolls up to the top of the potential due to the kinetic energy and the force induced by the potential and continues to accelerate the expansion of the Universe in the future.

\section{Conclusions}
\label{sec:4}

In this paper, we propose a pressure-parametric model for the investigation of the deviation from CC behavior of the dark sector accelerating the expansion of the Universe. Data from CMB anisotropies, BAO, SN Ia observations are employed to constrict the model parameters, coming out with the results that
$\Omega_{m0}=0.328 _{-0.008}^{+0.009}, \eta=-0.012_{-0.01}^{+0.01}, h=0.665_{-0.008}^{+0.008},\omega_{b}=0.0223_{-0.0002}^{+0.0002}$ at the 1$\sigma$ confidence level. We also consider the AIC to compare the new model and the $\Lambda$CDM model, which leads to a result of $\Delta AIC=0.83$.  To realize this model more physically, we reconstruct it with the quintessence and phantom scalar fields, and find out that although the model predicts a quintessence-induced acceleration of the Universe at past and present, when the scale factor $a>\frac{4}{3}$, dark energy turns into a phantom and continues to
accelerate the expansion of the Universe to its end.

\section*{Acknowledgments}
The paper is partially supported by the Natural Science Foundation of China.

\bibliographystyle{spphys}
\bibliography{pressure}

\end{document}